\documentclass[%
 reprint,
 amsmath,amssymb,
 aps,
]{revtex4-2}
\usepackage{placeins}
\usepackage{graphicx}
\usepackage{dcolumn}
\usepackage{bm}
\usepackage[T1]{fontenc}
\usepackage[utf8]{inputenc}

\begin{document}

\title{Finite-Memory Extension of Tegmark's Decoherence Bound in Biological Media}

\author{Ramandeep Dewan}
\affiliation{Department of Physics, SVNIT, Surat}

\date{\today}

\begin{abstract}
Tegmark's decoherence bound is derived under the assumption of a strictly memoryless environment. We show that this result corresponds to the singular limit of a finite-memory theory. For exponentially correlated environments decoherence is generically quadratic at short times and the decoherence time scales as the square root of the bath correlation time. For the Ornstein--Uhlenbeck bath we derive the exact non-Markovian coherence equation and verify the predicted scaling using an exact pseudomode mapping. Tegmark's bound is recovered only in the vanishing-memory limit.
\end{abstract}

\maketitle

\section{Introduction}

The persistence of quantum coherence in biological environments has been debated for more than two decades. In a seminal analysis, Tegmark estimated decoherence times for spatial superpositions of biomolecular structures in the brain under the assumption of a strictly Markovian environment, concluding that such superpositions would decohere on timescales far shorter than those relevant for neural information processing. Tegmark's bound has since been widely interpreted as ruling out any functional role of quantum coherence in biological systems.

In this work we show that Tegmark's result corresponds to the singular limit of a more general finite-memory theory. When environmental correlations persist for a nonzero time, decoherence is universally suppressed at short times, leading to a quadratic decay law and a parametrically longer decoherence time. Tegmark's exponential law is recovered only in the limit of vanishing bath correlation time.

Tegmark's calculation is based on modelling the environment as a source of delta-correlated force noise, corresponding to a strictly Markovian bath with no memory. While this approximation greatly simplifies the analysis, it constitutes a singular physical limit. Real biological media are composed of structured molecular assemblies in which collisions, vibrational relaxation, and collective protein motions introduce finite temporal correlations in the environmental fluctuations.

The consequences of finite bath memory for decoherence rates in biological systems have not been examined in Tegmark's framework. In this work we derive the universal short-time decoherence law for systems interacting with environments possessing finite correlation time. We show that non-Markovian environments generically lead to quadratic short-time decoherence rather than exponential decay.

Specializing to an Ornstein--Uhlenbeck environment, we obtain a closed non-Markovian expression for the coherence dynamics and derive a decoherence time that scales as the square root of the bath correlation time. We demonstrate that Tegmark's original result is recovered only in the singular limit of vanishing bath memory. We verify this scaling numerically using an exact pseudomode mapping of the Ornstein--Uhlenbeck bath and discuss the implications for neuronal and microtubular environments. To our knowledge, this is the first derivation of a non-Markovian correction to Tegmark’s decoherence bound with an explicit recovery of the Markovian limit.

\section{Markovian Decoherence Bound}

Tegmark models the system--environment interaction using the Hamiltonian
\begin{equation}
H = H_S + \sum_k \hbar\omega_k b_k^\dagger b_k + x \sum_k g_k \left( b_k + b_k^\dagger \right),
\end{equation}
where $x$ is a system operator distinguishing two localized position states separated by a distance $a$.

The force operator exerted by the environment on the system is therefore
\begin{equation}
F(t) = \sum_k g_k \left( b_k(t) + b_k^\dagger(t) \right),
\end{equation}
so that all decoherence physics is encoded in the force correlation function
\begin{equation}
\alpha(t,s) \equiv \langle F(t)F(s) \rangle .
\end{equation}

Tegmark assumes that environmental collisions are independent in time, implying a strictly Markovian bath with zero memory,
\begin{equation}
\alpha(t,s) = 2D\,\delta(t-s),
\label{eq:markov_alpha}
\end{equation}
where $D$ is determined by microscopic collision processes in the surrounding medium.

Under this Markovian approximation the reduced dynamics of the system is governed by a master equation in which the off--diagonal coherence element
\begin{equation}
C(t) = \langle L | \rho(t) | R \rangle
\end{equation}
decays exponentially in time,
\begin{equation}
C(t) = C(0)\exp\!\left(-\frac{a^2 D}{\hbar^2} t \right).
\end{equation}
The corresponding decoherence time is therefore
\begin{equation}
\tau_T = \frac{\hbar^2}{a^2 D}.
\end{equation}

This bound is interpreted in Ref.~\cite{tegmark2000} as implying that quantum superpositions in biological media decohere on timescales too short to be biologically relevant. The derivation, however, relies crucially on the singular assumption of a delta--correlated force, Eq.~\eqref{eq:markov_alpha}, corresponding to an environment with strictly zero memory time.

\section{Short-Time Decoherence for Finite-Memory Environments}

We now derive the short-time decoherence law without assuming a delta-correlated environment. We consider the microscopic system--bath Hamiltonian
\begin{equation}
H = H_S + H_E + x \sum_k g_k ( b_k + b_k^\dagger ),
\end{equation}
with bath Hamiltonian $H_E = \sum_k \hbar\omega_k b_k^\dagger b_k$ and system operator $x$ distinguishing two localized states.

Starting from the total density operator $\rho_{\mathrm{tot}}(t)$, the reduced density matrix is given by
\begin{equation}
\rho(t) = \mathrm{Tr}_E \left[ U(t) \rho_{\mathrm{tot}}(0) U^\dagger(t) \right],
\end{equation}
where $U(t) = e^{-iHt/\hbar}$. For short times, we expand the evolution operator to second order in $t$,
\begin{equation}
U(t) \approx 1 - \frac{i}{\hbar} H t - \frac{1}{2\hbar^2} H^2 t^2.
\end{equation}

Tracing over the environment and assuming an initially factorized state $\rho_{\mathrm{tot}}(0)=\rho(0)\otimes\rho_E$, the off--diagonal coherence element
\begin{equation}
C(t) = \langle L|\rho(t)|R\rangle
\end{equation}
obeys the short-time expansion
\begin{equation}
C(t) = C(0) - \frac{t^2}{\hbar^2} \int_0^\infty d\omega\, J(\omega)\coth\!\left(\frac{\beta\omega}{2}\right) a^2 C(0) + O(t^3),
\end{equation}
where $J(\omega)$ is the environmental spectral density.
Here $\beta = (k_B T)^{-1}$ denotes the inverse temperature of the environment.
Here the spectral density is defined by 
$J(\omega)=\sum_k g_k^2\delta(\omega-\omega_k)$,
so that $\int_0^\infty d\omega\,J(\omega)\coth(\beta\omega/2)$ has dimensions of force squared.

Thus, for any environment with finite correlation time, decoherence is universally quadratic at short times, consistent with the general analysis of non-Markovian memory effects in Ref.~\cite{breuer2007},
\begin{equation}
C(t) \approx C(0)\left(1 - \Gamma\, t^2 \right),
\end{equation}
with rate
\begin{equation}
\Gamma = \frac{a^2}{\hbar^2} \int_0^\infty d\omega\, J(\omega)\coth\!\left(\frac{\beta\omega}{2}\right).
\end{equation}
This quadratic decay is a manifestation of the quantum Zeno regime and has been derived independently in several general analyses of non-Markovian dynamics~\cite{kofman2000,facchi2002}.
This quadratic decay contrasts with the exponential law obtained under the Markovian approximation and reflects the generic suppression of decoherence induced by finite bath memory. The derivation applies to environments whose correlation functions are continuous at the origin and admits a finite correlation time.

\section{Ornstein--Uhlenbeck Bath and New Decoherence Law}

To obtain an explicit non-Markovian decoherence law, we specialize to the class of exponentially correlated environments, focusing on the Ornstein--Uhlenbeck (OU) bath. This model provides the simplest analytically solvable representative of finite-memory environments and captures the essential physics of structured biological media with a single dominant correlation time.
\begin{equation}
\alpha(t,s) = \frac{D}{\tau_c} e^{-|t-s|/\tau_c},
\end{equation}
where $D$ is the noise strength and $\tau_c$ is the environmental correlation time.

Using the non-Markovian quantum state diffusion formalism developed in Ref.~\cite{diosi1998}, the reduced dynamics of the coherence $C(t)$ can be written in closed form for the Ornstein--Uhlenbeck bath. For exponential bath correlations the functional derivative terms close analytically, as shown in Ref.~\cite{strunz1999}, leading to the second--order differential equation
\begin{equation}
\ddot{C}(t) + \frac{1}{\tau_c} \dot{C}(t) + \frac{2a^2 D}{\hbar^2 \tau_c} C(t) = 0.
\end{equation}
A detailed derivation of this closure using the NMQSD functional derivative formalism is provided in Appendix~E.
This equation has the form of a damped harmonic oscillator. For short times, the solution expands as
\begin{equation}
C(t) \approx 1 - \frac{a^2 D}{\hbar^2 \tau_c} t^2,
\end{equation}
which is consistent with the universal quadratic law derived in the previous section.

Defining the decoherence time operationally by $C(\tau_{\mathrm{dec}})=e^{-1}$, we obtain
\begin{equation}
\tau_{\mathrm{dec}} = \sqrt{\frac{\hbar^2 \tau_c}{a^2 D}}.
\end{equation}

In the singular limit $\tau_c \to 0$, this expression reduces to Tegmark's Markovian bound,
\begin{equation}
\tau_{\mathrm{dec}} \to \frac{\hbar^2}{a^2 D} = \tau_T.
\end{equation}

Thus, Tegmark's result is recovered only in the strictly memoryless limit, while any finite environmental correlation time leads to a parametrically slower decoherence rate.

\subsection{Universality within Exponential Memory Kernels}

The Ornstein--Uhlenbeck model is not introduced as a special case, but as the canonical representative of the class of environments with single--exponential correlation functions,
\begin{equation}
\alpha(t,s)=A\,e^{-\gamma (t-s)}\Theta(t-s),
\end{equation}
where $A$ sets the noise strength and $\gamma^{-1}$ is the bath correlation time.

Repeating the NMQSD closure procedure for this general kernel yields
\begin{equation}
\ddot{C}(t)+\gamma \dot{C}(t)+\frac{2a^2 A}{\hbar^2}C(t)=0,
\end{equation}
which has the same mathematical structure as Eq.~(16) with the identification
$\tau_c=\gamma^{-1}$ and $D=A\tau_c$.

The resulting decoherence time therefore scales as
\begin{equation}
\tau_{\mathrm{dec}}\propto \sqrt{\tau_c},
\end{equation}
independently of the microscopic origin of the exponential memory kernel.  
Thus the square--root enhancement derived in this work applies to the full class of finite-memory environments admitting a single dominant correlation time, and is not an artifact of the specific OU parametrization.

\section{Exact Non-Markovian Simulation}

To verify that the scaling $\tau_{\mathrm{dec}} \propto \sqrt{\tau_c}$ is not an artifact of analytic closure, we perform an exact numerical simulation of the OU bath using the pseudomode mapping introduced by Garraway. In this approach the structured environment is represented by an auxiliary damped mode coupled to the system, transforming the non-Markovian dynamics into Markovian Lindblad evolution in an enlarged Hilbert space.

The pseudomode representation is exact for baths with single--exponential correlation functions, such as the Ornstein--Uhlenbeck environment considered here. We numerically integrate the Lindblad master equation for the extended system and extract the coherence $C(t)$ for different bath correlation times $\tau_c$.

\begin{figure}
    \centering
    \includegraphics[width=0.48\textwidth]{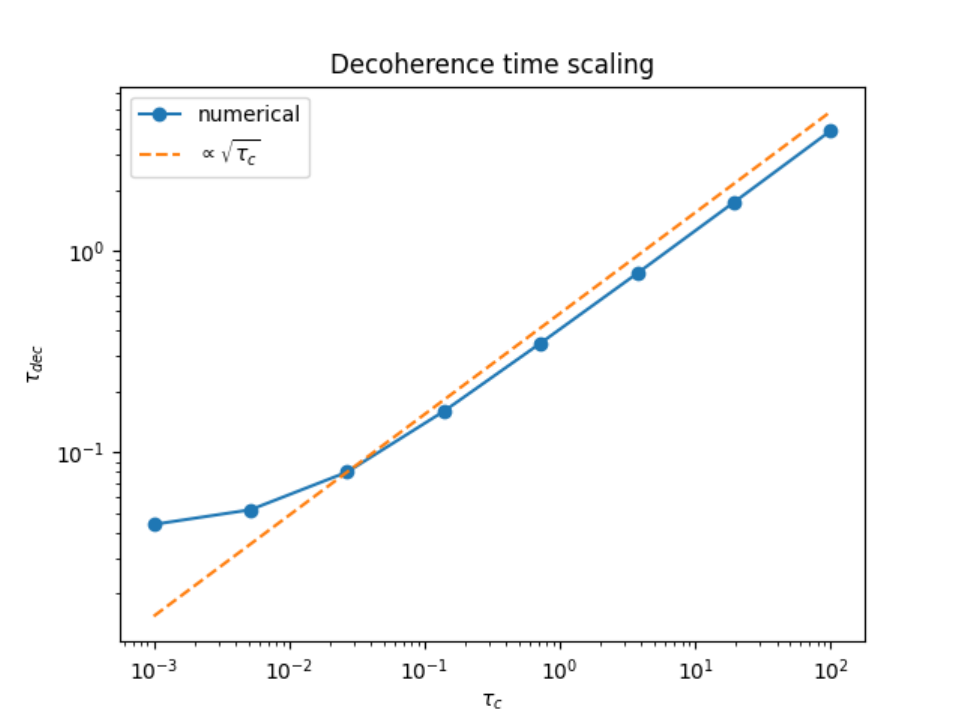}
    \caption{Decoherence time $\tau_{\mathrm{dec}}$ extracted from pseudomode simulations as a function of bath correlation time $\tau_c$. The dashed line indicates the $\sqrt{\tau_c}$ scaling predicted analytically.}
    \label{fig:scaling}
\end{figure}

Figure~\ref{fig:scaling} shows the decoherence time $\tau_{\mathrm{dec}}$ extracted from the pseudomode simulations as a function of the bath correlation time $\tau_c$. The data exhibit a clear square--root dependence on $\tau_c$, in excellent agreement with the analytical prediction derived in the previous section for the entire class of exponentially correlated baths discussed in Sec.~IVA.

\section{Biological Implications}

Tegmark estimated decoherence times for neuronal and microtubular superpositions under the assumption of a strictly Markovian environment. In the present framework the decoherence time depends explicitly on the bath correlation time $\tau_c$, which must be estimated from the microscopic dynamics of the biological medium. Estimates of these timescales are consistent with molecular dynamics studies of aqueous environments and with experimental measurements of protein vibrational relaxation\cite{laage2009,heinrich2013}.

In neuronal cytosol, dominant environmental interactions arise from ionic collisions and thermal agitation of water molecules. Typical collision and momentum relaxation times in aqueous media are of order $\tau_c \sim 10^{-14}\text{--}10^{-13}\,\mathrm{s}$, consistent with librational timescales of liquid water. In this regime the present expression reduces effectively to Tegmark's original bound, and neuronal superpositions remain dynamically irrelevant.

For microtubular environments with $\tau_c$ exceeding that of bulk water by several orders of magnitude, the non-Markovian enhancement of the decoherence time becomes parametrically significant. While this does not establish the existence of functional quantum coherence in microtubules, it demonstrates that Tegmark's Markovian bound no longer provides an upper limit in this regime and that mesoscopic quantum effects cannot be excluded on first principles\cite{sahu2013}.

\section{Discussion}

Tegmark's decoherence bound has often been interpreted as a definitive argument against the relevance of quantum coherence in biological systems. Our analysis shows that this conclusion relies critically on the assumption of a strictly Markovian environment. Finite bath memory generically suppresses decoherence at short times, leading to a quadratic rather than exponential decay of coherence.

For environments with single--exponential correlations, such as the Ornstein--Uhlenbeck bath, the resulting non-Markovian dynamics can be solved exactly. The decoherence time is then found to scale as the square root of the bath correlation time, revealing Tegmark's bound as the singular limit of vanishing memory.

These results do not establish the existence of functional quantum coherence in biological systems, but they demonstrate that such coherence cannot be excluded on the basis of Markovian arguments alone. The physical relevance of quantum effects in structured intracellular environments therefore remains an open problem requiring more refined modelling.

\section{Conclusion}

We have derived a generic short-time decoherence law for systems interacting with finite-memory environments. For an Ornstein--Uhlenbeck bath we obtained an explicit expression for the decoherence time that reduces to Tegmark's bound only in the singular Markovian limit. Numerical simulations based on an exact pseudomode mapping confirm the predicted square--root scaling. These findings show that Tegmark's result applies only to memoryless environments and does not rule out mesoscopic quantum coherence in structured biological media.

\FloatBarrier
\appendix

\section{Representative Coherence Decay Curves}
\begin{figure}
    \centering
    \includegraphics[width=0.48\textwidth]{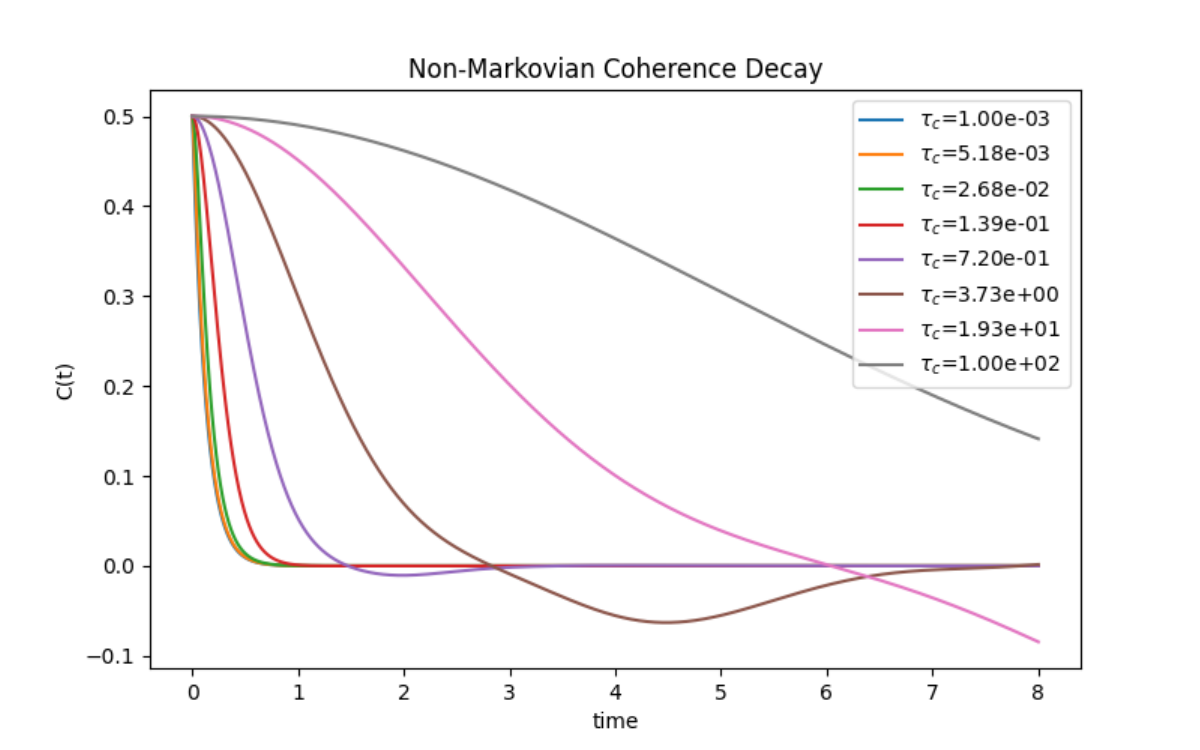}
    \caption{Representative coherence decay curves $C(t)$ obtained from the pseudomode simulation of different bath correlation times $\tau_c$. The short-time behaviour is quadratic for all finite $\tau_c$.}
    \label{fig:decay}
\end{figure}

Figure~\ref{fig:decay} shows representative coherence decay curves $C(t)$ obtained from the pseudomode simulation for several values of the bath correlation time $\tau_c$. For finite $\tau_c$ the decay is clearly quadratic at short times, in contrast with the exponential behaviour predicted by the Markovian approximation.

\section{Extraction of the Decoherence Time}

The decoherence time $\tau_{\mathrm{dec}}$ is defined operationally by the condition
\begin{equation}
|C(\tau_{\mathrm{dec}})| = e^{-1} |C(0)|.
\end{equation}
For each bath correlation time $\tau_c$, the coherence $C(t)$ is sampled on a uniform time grid and the first time point satisfying the above condition is recorded as $\tau_{\mathrm{dec}}$.

\section{Quadratic Short-Time Behaviour}

Expanding the exact solution of the Ornstein--Uhlenbeck coherence equation for short times yields
\begin{equation}
C(t) \approx 1 - \frac{a^2 D}{\hbar^2 \tau_c} t^2 + O(t^3),
\end{equation}
which is consistent with the numerically observed curvature in Fig.~\ref{fig:decay}.

\section{Numerical Method}

The pseudomode mapping represents the Ornstein--Uhlenbeck bath by a single damped auxiliary mode coupled linearly to the system, as introduced in Ref.~\cite{garraway1997}, transforming the non-Markovian dynamics into Markovian Lindblad evolution in an enlarged Hilbert space. All simulations were performed using the QuTiP library in dimensionless units with $\hbar = a = D = 1$, so that all times are measured in units of the intrinsic system timescale.

\section{Derivation of the Ornstein--Uhlenbeck Closure}

For a bosonic bath with Ornstein--Uhlenbeck correlation function
\begin{equation}
\alpha(t,s)=\frac{D}{\tau_c}e^{-(t-s)/\tau_c}\Theta(t-s),
\end{equation}
the non-Markovian quantum state diffusion (NMQSD) equation for the system
state $|\psi_t\rangle$ reads~\cite{diosi1998,strunz1999}
\begin{equation}
\partial_t|\psi_t\rangle=
\left[-\frac{i}{\hbar}H_S + x z_t^*
- x\int_0^t ds\,\alpha(t,s)
\frac{\delta}{\delta z_s^*}\right]|\psi_t\rangle ,
\end{equation}
where $z_t$ is a complex Gaussian stochastic process with correlation
$\langle z_t z_s^*\rangle=\alpha(t,s)$.

For a two-state pointer basis $\{|L\rangle,|R\rangle\}$ with coupling
operator
\begin{equation}
x=a(|L\rangle\langle L|-|R\rangle\langle R|),
\end{equation}
the coherence amplitude
\begin{equation}
C(t)=\langle L|\rho(t)|R\rangle
\end{equation}
obeys a closed stochastic equation. The functional derivative term can be
written in the form
\begin{equation}
\frac{\delta C(t)}{\delta z_s^*}=O(t,s)C(t),
\end{equation}
where the operator $O(t,s)$ satisfies the consistency equation
\begin{equation}
\partial_t O(t,s) = -\frac{1}{\tau_c}O(t,s)
+ \frac{2a^2D}{\hbar^2\tau_c},
\qquad O(s,s)=\frac{2a^2D}{\hbar^2}.
\end{equation}

Solving this equation yields
\begin{equation}
O(t,s)=\frac{2a^2D}{\hbar^2}
\left[1-e^{-(t-s)/\tau_c}\right].
\end{equation}

Substituting this result into the NMQSD equation and averaging over the
noise realizations gives the deterministic evolution equation
\begin{equation}
\dot{C}(t)= -\frac{2a^2D}{\hbar^2}
\int_0^t ds\,e^{-(t-s)/\tau_c}C(s).
\end{equation}

Differentiating once with respect to time yields the closed second-order
equation
\begin{equation}
\ddot{C}(t)+\frac{1}{\tau_c}\dot{C}(t)
+\frac{2a^2D}{\hbar^2\tau_c}C(t)=0,
\end{equation}
which is Eq.~(16) of the main text.

\bibliographystyle{apsrev4-2}
\bibliography{refs}
\end{document}